\documentclass[prb,twocolumn,showpacs,floatfix,amsfonts]{revtex4}
\usepackage{epsfig}
\begin{document}
\title{Influence of temperature on the nodal properties
of the longitudinal thermal conductivity of
YBa$_2$Cu$_3$O$_{7-x}$}
\author{Roberto Oca\~{n}a and Pablo Esquinazi}
\address{Abteilung  Supraleitung und Magnetismus, Institut f\"ur
Experimentelle Physik II, Universit\"at Leipzig, Linn{\'e}str.~5,
D-04103 Leipzig, Germany}
\begin{abstract}
The angle dependence at different temperatures of the longitudinal
thermal conductivity $\kappa_{xx}(\theta)$ in the presence of a
planar magnetic field is presented. In order to study the
influence of the gap symmetry on the thermal transport, angular
scans were measured up to a few Kelvin below the critical
temperature $T_c$. We found that the four-fold oscillation of
$\kappa_{xx}(\theta)$ vanishes at $T > 20~$K and transforms into a
one-fold oscillation with maximum conductivity for a field of 8~T
applied parallel to the heat current. Nevertheless, the results
indicate that the $d-$wave pairing symmetry remains the main
pairing symmetry of the order parameter up to $T_c$. Numerical
results of the thermal conductivity using an Andreev reflection
model for the scattering of quasiparticles by supercurrents under
the assumption of $d-$wave symmetry provide a semiquantitative
description of the overall results.
\end{abstract}
\maketitle
\section{Introduction}
The order parameter in high-temperature superconductors (HTS) has
been proved to have mainly a $d_{x^2-y^2}$ pairing
symmetry.\cite{tsu,aub,sal,oca3} In particular,  try-crystal
phase-sensitive measurements \cite{tsu} have determined the
presence of a predominant $d_{x^2-y^2}-$gap symmetry up to the
superconducting critical temperature $T_c$ over, if any, other
minor components smaller than 5\% of that with the
$d_{x^2-y^2}-$symmetry. Thermal transport measurements have also
shown the predominance of this symmetry for the gap function. In
particular, when a magnetic field is rotated parallel to the
CuO$_2$ planes, the longitudinal thermal conductivity shows a
fourfold oscillation which can be explained in terms of both
Andreev scattering of quasiparticles by vortices (AS) and  Doppler
shift (DS) in the energy spectrum of the quasiparticles if one
takes a $d_{x^2-y^2}-$gap into account.\cite{aub,sal,oca3}
Nevertheless, the fourfold oscillation in the thermal conductivity
has been resolved up to $\sim 15$ K. Above this temperature, no
direct evidence for this type of symmetry from this kind of
measurements has been published. Furthermore, deviations from the
expected angular pattern within a pure $d_{x^2-y^2}-$symmetry have
been attributed to the effect of pinning of vortices.\cite{aub}

The variation of the thermal conductivity  as a function of angle $\theta$
between the heat current and the magnetic field applied parallel to the
CuO$_2$ planes depends mainly on the heat transport by quasiparticles,
their interaction with the supercurrents (vortices) and the symmetry of
the order parameter.\cite{aub,sal,oca3} An angular pattern showing
properties of the order parameter symmetry can only be achieved when the
temperature is low enough so that the quasiparticle momentum is close to
the nodal directions of the gap. Otherwise, thermal activation would also
induce quasiparticles at different orientations from those of the nodes
and hence, the sensitivity of the probe to measure gap characteristics
will be reduced. The first question we address in this paper is related to
the temperature range at which the nodal characteristics of the gap are
directly observable by thermal conductivity. The second question we would
like to clarify in this paper is whether the thermal activation of
quasiparticles with increasing temperature, treated phenomenologically
within a Fermi liquid approximation for thermal transport including the
$d_{x^2-y^2}-$gap function, can explain the experimental results in the
whole temperature range, in particular the change of symmetry of
$\kappa_{xx}(\theta)$ as a function of temperature. In this paper, we
calculate numerically the thermal conductivity at different angles and
temperatures at fixed magnetic field, assuming an Andreev reflection model
for the scattering of quasiparticles by supercurrents, originally proposed
by Yu et al. \cite{yu,yucomment,yureply} within the  two dimensional BRT
expression \cite{bar} for the thermal conductivity, and compare it to the
experimental data.

Angle scans in a magnetic field applied parallel to the CuO$_2$
planes were performed in order to measure the angular variation of
the longitudinal thermal conductivity $\kappa_{xx}(\theta)$ in two
single crystals of YBa$_2$Cu$_3$O$_{7-x}$ high-temperature
superconductor. The overall results agree with the theoretical
model and confirm the predominance of the $d_{x^2-y^2}-$gap up to
$T_c$. The measurements provide also new results that improve our
knowledge of the thermal transport at temperatures at which the
nodal properties of the gap are not directly observable.

Following the experimental and sample details of the next section,
we present in Sec.~\ref{res} the main experimental results. In
Sec.~\ref{comparison} we describe the used model and compare it
with the experimental data. A brief summary is given in
Sec.~\ref{sum}.
\section{Experimental and Sample details}
\label{exp} In order to rule out effects concerning shape and
structure characteristics of the crystal we have used two
different samples of YBa$_2$Cu$_3$O$_{7-x}$ (YBCO): a twinned
single crystal  with dimensions (length $\times$ width $\times$
thickness) $0.83 \times 0.6 \times 0.045~$mm$^3$ and critical
temperature $T_c =93.4~$K previously studied in
Refs.~\cite{tal,oca1,oca2,oca3} and an untwinned single crystal
with dimensions $2.02 \times 0.68 \times 0.14~$mm$^3$ and $T_c =
88~$K.\cite{oca2} For the measurement of the thermal conductivity,
a heat current $J$ was applied along the longest axis of the
crystal studied. In the untwinned sample, J was parallel to the
$a$-axis and in the twinned crystal was parallel to the $a/b$-axes
(twin planes oriented along (110)). In both cases the position of
the lattice axis with respect to the crystal axis was determined
using polarized light microscopy and X-ray diffraction.

\begin{figure}
\begin{center}
\centerline{\psfig{file=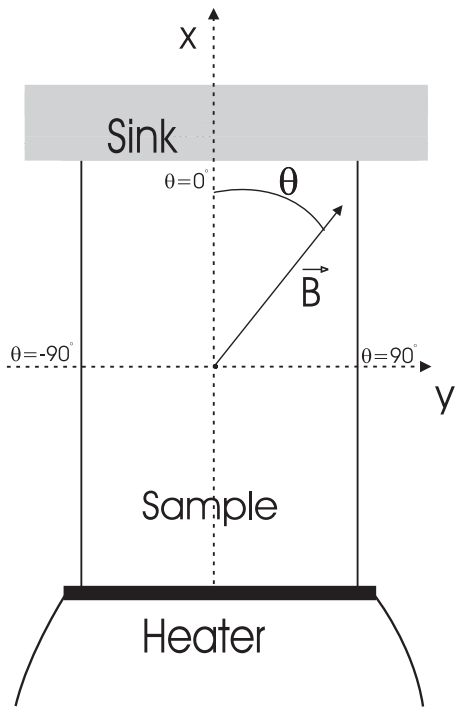,width=6.0cm}} 
\end{center}
\caption[*]{Top view of the sample arrangement and the definition of the
angle $\theta$.} \label{esquema}
\end{figure}

The longitudinal temperature gradient $(\nabla_x T)$ was measured
using previously calibrated chromel-constantan (type E)
thermocouples \cite{iny} and a dc picovoltmeter. Special efforts
were made in order to minimize the misalignment of the plane of
rotation of the magnetic field applied perpendicular to the c-axis
with the CuO$_2$ planes of the sample. This misalignment was
minimized step by step, measuring the angle dependence of
$\kappa_{xx}(\theta)$ until a satisfactory symmetrical curve was
obtained. In this way, we estimate a misalignment smaller than
$0.5^0$. An in-situ rotation system enabled  measurement of the
thermal conductivity as a function of the angle $\theta$ defined
between the applied field and the heat flow direction along $+
\widehat{x}$, see Fig.~\ref{esquema}. For more details on the
experimental arrangement see Ref.~\cite{oca3}.

As pointed out by Aubin et al.~\cite{aub} and observed in
Refs.~\cite{tal,oca3}, the effect of the pinning of vortices plays
an important role in determining the correct angle pattern in this
kind of measurement. In fact, when the angle of the magnetic field
is changed, a non uniform vortex distribution due to pinning
forces may appear. As argued in Ref.~\cite{oca3}, the pinning of
the Josephson-like vortices parallel to the planes is strongly
affected by vortices perpendicular to the planes which may appear
due to the misalignment of the crystal axes with respect to the
applied magnetic field. In this situation, even hysteresis in the
angular patterns of the thermal conductivity can be
measured.~\cite{aub} We note that pinning of vortices is
influenced by the distribution of the oxygen vacancies in the
sample as well as by defects and impurity centers. Thus, in order
to rule out the influence of this effect in the measurements, a
field-cooled procedure have to be used. This procedure consists of
two steps. In the first step the sample is driven into the normal
state by heating to a few Kelvin above $T_c$ and the angle is
changed. Secondly, it is cooled down to the desired temperature at
constant field.
\begin{figure}
\begin{center}
\centerline{\psfig{file=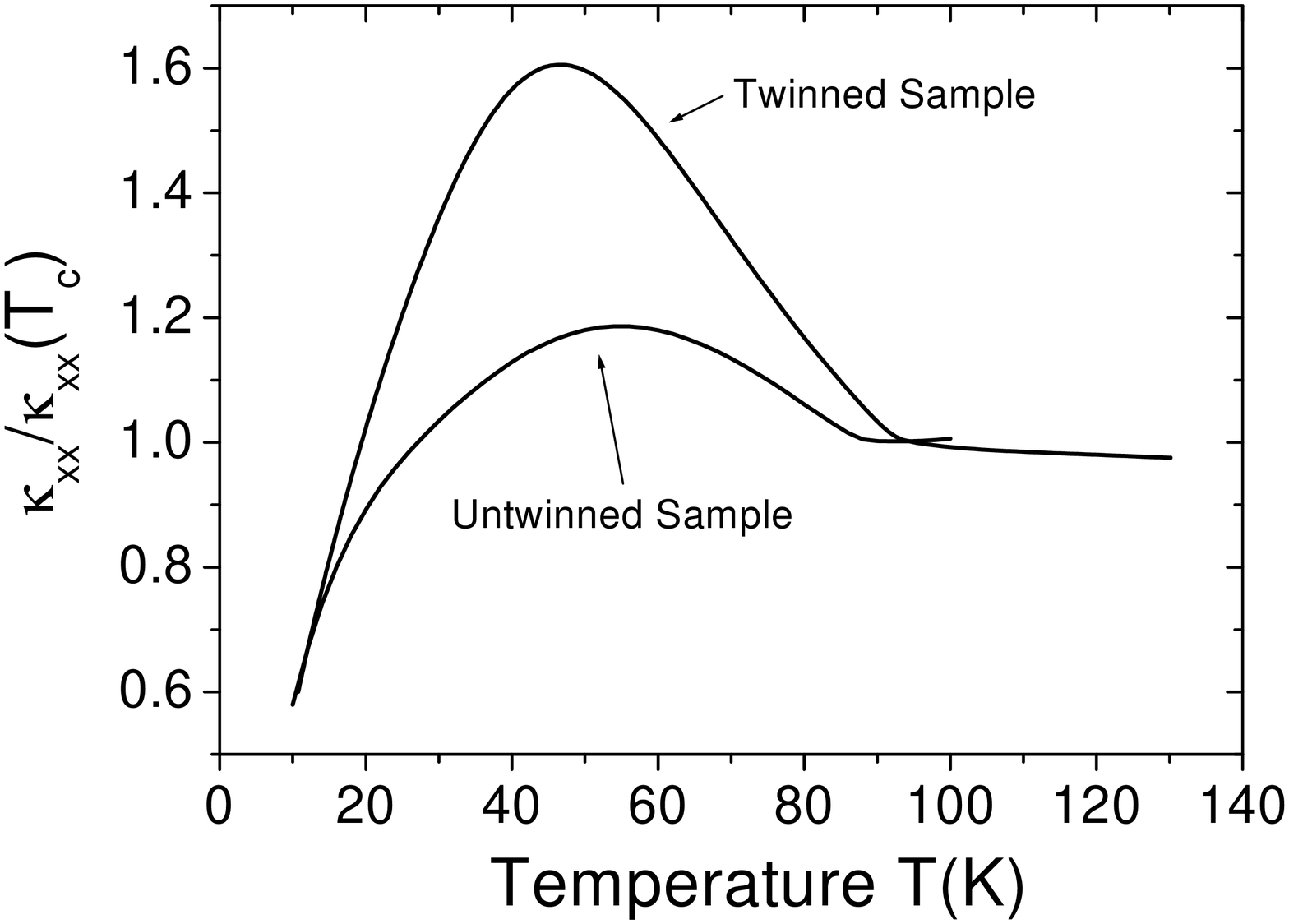,width=\columnwidth}}
\end{center}
\caption[*]{Temperature dependence of the longitudinal thermal
conductivity $\kappa_{xx}$ for the twinned and untwinned crystals.
The curves are normalized by the value of the thermal conductivity
at $T_c$. Within $\sim 30\%$ error the absolute value of the
thermal conductivity at $T_c$ for the untwinned (twinned) crystal
is $\kappa_{xx}(T_c)\simeq 12 (8)~$W/Km.} \label{fig1}
\end{figure}

In Fig.~\ref{fig1} we show the longitudinal thermal conductivity
as a function of the temperature in both crystals at zero magnetic
field. As argued by many authors (see for example
Refs.~\cite{hir2,hir1,kub1}), the observed behavior in
Fig.~\ref{fig1} provides qualitative information about the quality
of the sample. The height of the peak observed in the temperature
dependence of the thermal conductivity is related to the relative
contributions between the density of impurity scattering centers
and the strength of the inelastic electron-electron scattering.
Crystals showing large peaks may have a small amount of impurity
scattering centers \cite{zeiniepj} and/or larger
quasiparticle-related inelastic contribution. The origin of the
peak is explained in terms of a competition between the decrease
of the inelastic scattering rate~\cite{bonn} and the decrease of
the population of quasiparticles as temperature
decreases.\cite{hir1,houssa1,houssa2} Taking this into account we
might conclude that the untwinned crystal has a larger impurity
scattering and a smaller quasiparticle-related inelastic
scattering since its peak is considerably smaller than that of the
twinned sample. This is supported by its reduced T$_c$ with
respect to the twinned sample (which indicates a larger density of
oxygen vacancies). Since the thermal conductivity at $T_c$ appears
to be larger for the untwinned sample, we conclude that there
should be a substantial reduction of the inelastic scattering. As
we show below, this makes the measurement of electronic properties
related to the gap symmetry more difficult.
\section{Experimental Results and Discussion}\label{res}
\begin{figure}\vspace{-0.5cm}
\begin{center}
\centerline{\psfig{file=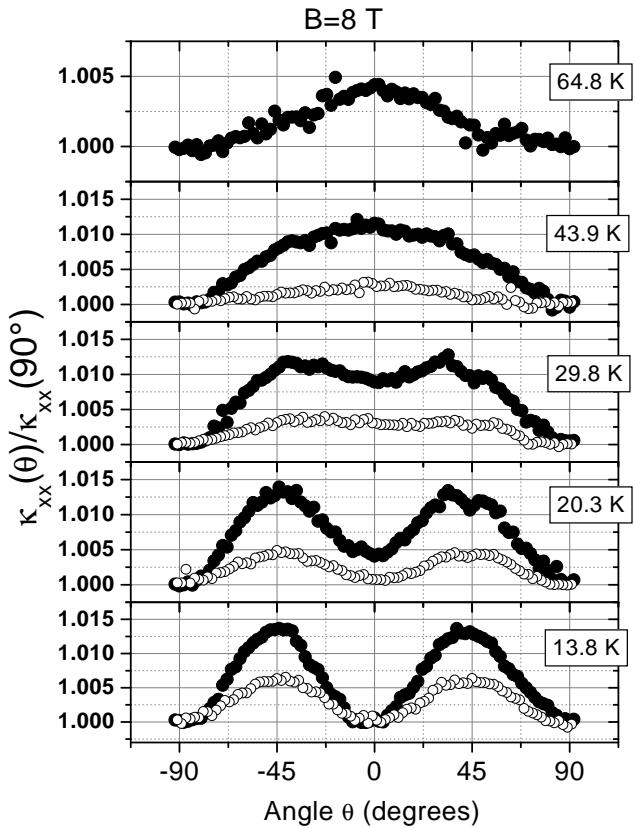,width=\columnwidth}}
\vspace{-1.cm}
\end{center}
\caption[*]{Angle dependence  of the longitudinal thermal
conductivity at different temperatures for the twinned sample. The
results are normalized to the value of the thermal conductivity at
$\theta = 90^{\circ}$, when the field was applied perpendicular to
the heat current. Solid circles are taken in a magnetic field
strength of 8 T and open circles at 3 T.} \label{fig2}
\end{figure}
\begin{figure}
\vspace{-0.5cm}
\centerline{\psfig{file=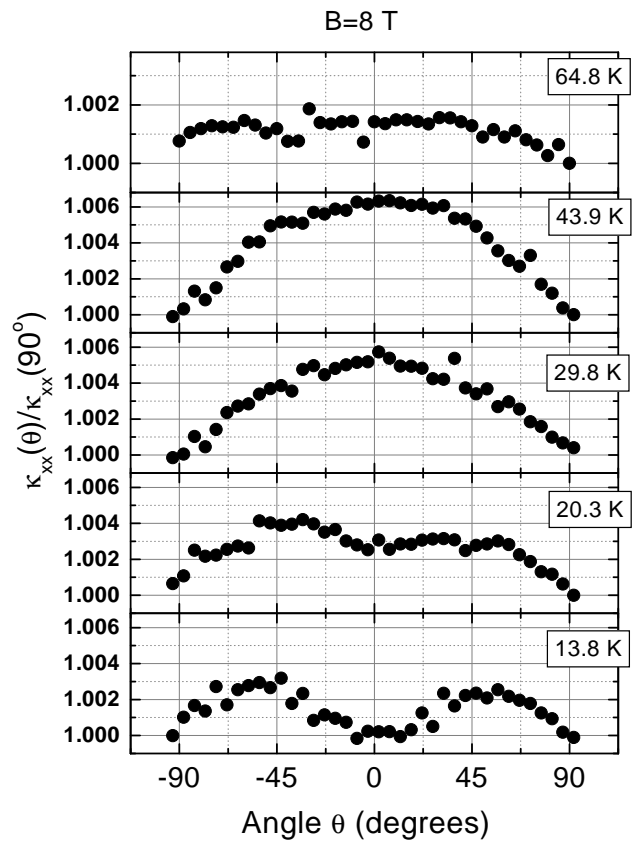,width=\columnwidth}}
\vspace{-1.0cm} \caption[*]{Angle dependence patterns of the
longitudinal thermal conductivity at different temperatures and 8
T for the untwinned sample. The results are normalized to the
value of the thermal conductivity when the field was applied
perpendicular to the heat current, at $\theta = 90^{\circ}$. }
\label{fig3}
\end{figure}

The angular patterns of the longitudinal thermal conductivity
$\kappa_{xx}$ at different temperatures for the twinned and
untwinned  crystals are shown in Figs.~\ref{fig2} and \ref{fig3}
respectively. At low enough temperatures ($T < 20 K$), a magnetic
field parallel to the CuO$_2$ planes of the sample produces a
fourfold oscillation in the longitudinal thermal conductivity as
the field is rotated from $\theta = 90^{\circ }$ to $\theta =
-90^{\circ }$. We note that at both $\theta = 90^{\circ }$ and
$\theta = -90^{\circ }$ the magnetic field is perpendicular to the
heat flow. At $\theta =0^{\circ }$ the field is  parallel to the
heat flow (see Fig.~\ref{esquema}). As noted in Refs.~\cite{aub}
and~\cite{oca3} the angular patterns are not free of the phononic
contribution to the total thermal conductivity. However, this
contribution seems to be constant under variations of the magnetic
field orientations for two reasons. 1) For fields parallel to the
CuO$_2$ planes, vortices are unlikely to have a normal core
(Josephson vortices), making a change of the phonon attenuation
with field and angle unlikely.  2) There is no experimental
evidence that the phononic contribution does change substantially
with field. The similarity between the temperature dependencies of
$\kappa_{xx}$ and $\kappa_{xy}$ below $T_c$ where a large change
in the quasiparticle density ocurrs\cite{oca1}, indicates that the
phonon-electron interaction does not play a main role in the
temperature dependence of $\kappa$. The electronic contribution is
entirely responsible for the magnetic field dependence, revealed
in both thermal conductivity~\cite{yu2,zeini,kris2,oca1} and in
microwave measurements~\cite{bonn}, and indicates against any
large phonon-electron scattering. In fact, no remarkable
phonon-electron scattering have been addressed in those
experiments where the measured magnetic field dependencies are
larger than the oscillation amplitudes measured in this paper.
Therefore, measurable contributions of the phonons via
phonon-electron scattering are more unlikely to occur in the
angular patterns. This characteristic makes the angular profiles
of the thermal conductivity suitable to be compared with
electronic models by using the quantity
$\kappa_{xx}(\theta)-\kappa_{xx}(90^{\circ})$ where the phononic
contribution is subtracted\cite{oca3}.

As pointed out in Ref.~\cite{oca3}, the variation of the thermal
conductivity in Figs.~\ref{fig2} and~\ref{fig3} can be explained
with a model involving a Doppler shift (DS) in the energy spectrum
of the quasiparticles  along with an accurate inclusion of the
impurity scattering \cite{won,won2,hir1,kub1} and/or assuming
Andreev scattering (AS) of quasiparticles by vortices\cite{yu}.
Briefly, in the mixed state the quasiparticles are in the presence
of a phase gradient produced by the superfluid flow of the
vortices. Thus, as viewed from the laboratory frame they
experience a Doppler shift in their energy spectrum given by the
scalar product of the momentum and superfluid velocity,
\textbf{p}$\cdot$\textbf{v}$_s$. When a quasiparticle of momentum
\textbf{p} is moving parallel to the magnetic field, the product
is zero and hence, no DS occurs. Thus, the angular characteristic
of this effect is to produce an excess of quasiparticles in the
direction perpendicular to the field and thereby, reducing the
``local" thermal resistance at this orientation
\cite{won,won2,hir1}. When the field is placed parallel to the
heat current, the Doppler shift is the same for both nodal
directions at $\theta = 45^{\circ }$ and $\theta = -45^{\circ }$.
Therefore and since thermal resistances must be added in parallel,
a simple picture in which there are only quasiparticles at the
nodes would produce a fourfold oscillation in the thermal
conductivity with opposite sign to that observed in the
measurements. However, as pointed out in
Refs.~\cite{kub1,hir1,won2}, the DS affects both the carrier
density as well as the scattering rate of quasiparticles, and
since at high enough temperatures the latter dominates, the
quasiparticles move more easily parallel to the magnetic field.
Therefore, the sign of the fourfold oscillation observed in the
experiments is also recovered within this picture.

In the AS mechanism a phase gradient, namely, the superfluid flow
surrounding the vortices may induce Andreev reflection of the
quasiparticles \cite{andi}. Thus, the DS in the energy spectrum
\textbf{p}$\cdot$\textbf{v}$_s$ is implicitly taken into account in the AS
picture. As viewed from the laboratory frame, when the quasiparticle
energy equals the value of the gap, then the quasiparticle is transformed
into a quasihole reversing its velocity and hence, decreasing its
contribution to the thermal conductivity. Thus, a quasiparticle with
momentum \textbf{p} parallel to the magnetic field does not experience a
DS and hence, no Andreev reflection can take place. The field acts as a
filter~\cite{mat} for the quasiparticle that contributes to reduce the
total temperature gradient. Note that the AS picture gives rise to a
fourfold oscillation in the thermal conductivity as the magnetic field is
rotated parallel to the CuO$_2$ planes without more considerations.

Then, qualitatively both AS and DS can explain in principle the angle
profiles observed at low temperatures. However as pointed out in
Ref.~\cite{oca3}, neither AS nor DS alone can explain the magnetic field
dependence of the oscillation amplitude in the whole field range
$0~$T$~\le B \le 9~$T. Therefore, a more realistic picture of the thermal
transport should take both into account. As argued in Ref.~\cite{vek2},
this scenario produces different regimes influenced by the predominance of
either the DS or the AS in the magnetic field dependence. Thus, at
constant temperature the strength of the magnetic field becomes the
parameter that changes the regime. At low magnetic fields the DS dominates
and the AS dominates at high fields \cite{vek2,oca3}. In both cases, an
increase of the oscillation amplitudes with increasing magnetic field
strength is predicted. In Fig.~\ref{fig2}  the oscillation amplitudes of
the thermal conductivity at 3~T and 8~T for different temperatures in the
twinned crystal are shown.

As the temperature increases the fourfold oscillation is no longer
observable, see Figs.~\ref{fig2} and \ref{fig3}. This fact can be
understood if one takes into account the thermal activation of the
quasiparticles at different orientations from those of the nodes
of the order parameter. In other words, if we suppose that the
same processes that govern the thermal transport at low
temperatures (AS and DS) are responsible for the high temperature
behavior as well, we have to conclude that an increasing number of
carriers appears in the direction of the heat current as the
temperature is raised. As we shall see below from the numerical
results using the AS model, this is, in fact, what takes place as
the temperature is increased. However, although the observed angle
profiles can be well described by the numerical analysis, the
amplitude of the oscillations are smaller than the simulation
results at $T > 30~$K. As argued for the peak of the curves in
Fig.~\ref{fig1} and shown in the following section, this
discrepancy can be solved by the inclusion of an inelastic
scattering into the calculations.\cite{hir2,bonn}

In Fig.~\ref{fig3} we show the angle dependence of the untwinned
crystal at 8 T. Similar angle profiles as for the twinned sample
have been found, but the amplitude of the oscillation is
considerably smaller. This can be explained in terms of a larger
concentration of impurity scatterers in the untwinned sample, as
discussed above and in Ref.~\cite{oca2}. In fact, if the impurity
scattering rate is sufficiently large, the relative weight of the
directionality of the AS and DS mechanisms to
$\kappa_{xx}(\theta)$ weakens. The influence of the impurity
scattering has been considered in the original models by inclusion
of an impurity rate.\cite{won,won2,hir1,yu}

We note also that the symmetry of the $\kappa_{xx}(\theta)-$curves
differs slightly for the twinned and untwinned samples at the same
absolute or reduced temperature. The effect of impurity scatterers
can be accounted for by an impurity dependent gap parameter as
done in Ref.~\cite{won}. In general, however, it can be viewed as
an effect related to the ratio of the modulus of the gap and the
density of states of the quasiparticles at a temperature T and
momentum {\bf p}. Therefore, since our twinned sample has a larger
critical temperature $T_c$ than our untwinned sample, the oxygen
deficiency in the latter could be also responsible for a reduced
density of states, and therefore, for a different angle patterns
respect to the sample with a higher $T_c$.
\begin{figure}
\vspace{-0.75cm}
\epsfig{file=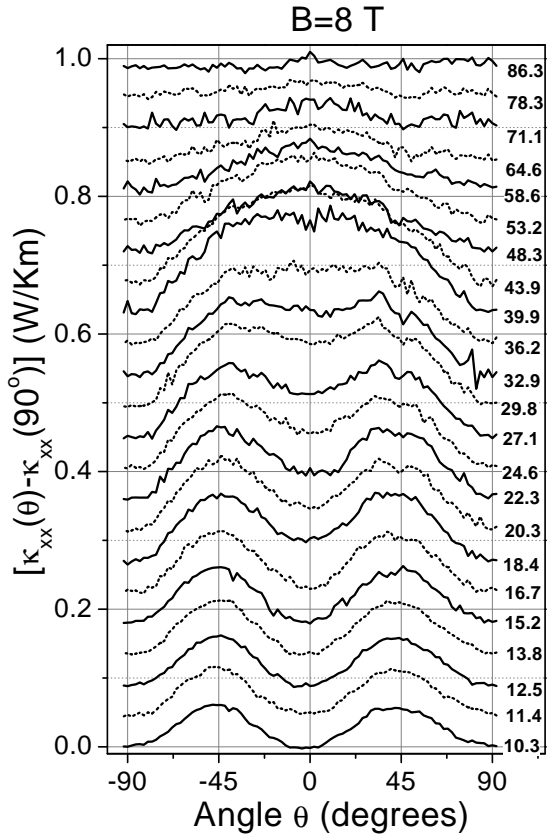,width=8.0cm}
\caption[*]{Angle dependence patterns of the longitudinal thermal
conductivity at different temperatures and at 8~T for the twinned
sample. The value of the thermal conductivity at $\theta =
90^{\circ}$ was subtracted to facilitate the comparison with
theory. An offset has been added to every curve in order to
distinguish the change in the symmetry as the temperature
increases.} \label{fig4}
\end{figure}

Figure~\ref{fig4} shows the complete angle patterns measured in
the twinned crystal at 8~T. At $T > 78$ K no clear change in the
thermal conductivity when the field is rotated parallel to the
CuO$_2$ planes is observed. This can be explained by taking into
account the temperature dependence of the gap, which vanishes at
$T_c$, and the relative increase of the inelastic scattering rate.
\section{Comparison with the two-dimensional thermal transport
theory}\label{comparison}

To compare the experimental results with theory we take recently
published results of the longitudinal and transverse thermal
conductivity\cite{oca3} into account  that indicate that at high
fields ($B > 2~$T), applied parallel to the planes, the main
scattering mechanism for the quasiparticles appears to be the
Andreev reflection by vortex supercurrents.\cite{andi} Therefore,
we use the formulation proposed by Yu et al.\cite{yu} using a two
dimensional model of the BRT expression for the thermal
conductivity,\cite{bar} which has been useful in the
interpretation and discussion of previous
results.\cite{aub,oca2,oca3} Under this model the longitudinal
thermal conductivity can be written as
\begin{equation}
\kappa^{el}_{xx} = \frac{1}{2\pi^2ck_BT^2\hbar^2}
\int^\infty_{p_F} {\rm d}^2p \frac{v_{gx}v_{gx} E_{\bf
p}^2}{\Gamma({\bf B},{\bf p},T)} {\rm sech}^2 \left(\frac{E_{\bf
p}}{2 k_B T} \right) \,, \label{brd}
\end{equation}
where $v_{gx}$ is the the x-axis component of the group velocity,
and $E_{\bf p}$ is the quasiparticle energy. For this energy we
use a free quasiparticle model $E^2_{\bf p} = ({\bf p}^2/2m_{\rm
eff} - \mu)^2 + \Delta^2({\bf p},T)$ where $m_{\rm eff}$ is the
effective mass of the quasiparticle. $\Gamma ({\bf B},{\bf p},T)$
may be taken as a relaxation rate given by the sum of the
following scattering mechanisms acting in series: scattering of QP
by impurities $\Gamma_{\rm imp}({\bf p})$, by phonons $\Gamma_{\rm
ph}({\bf p},T)$,\cite{houssa1,houssa2,houssa3} by quasiparticles
$\Gamma_{\rm qp}({\bf B},{\bf p},T)$\cite{hir2,quin} and AS by
vortex supercurrents $\Gamma_{v}({\bf B},{\bf p},T)$.\cite{yu} The
expression for this scattering rate according to the model for the
AS, proposed by Yu et al. \cite{yu}, is given by
\begin{equation}
\Gamma_{v} ({\bf B},{\bf p},T)=\Gamma^0_{v}\exp \left\{
\frac{-m^2a_\upsilon ^2\left[ E_p-\left| \Delta ({\bf p},T)\right|
\right] ^2}{p_F^2\hbar ^2\ln (a_\upsilon /\xi _0)\sin ^2\psi ({\bf
p})}\right\}\,,
\end{equation}
where $a_\upsilon$ is the intervortex spacing given in this model
by
\begin{equation}
a^2_\upsilon = \frac{p_F^2\hbar ^2}{\pi\Delta^2_0\gamma m^2_e} \,
\frac{B^{ab}_{c2}}{B}\,.\label{spacing}
\end{equation}
We use the BCS-like parameterization for the temperature
dependence of the gap amplitude
\begin{equation}
\Delta (T) = \Delta_0 \tanh\left( 2.2\,\sqrt {{\frac {{\it
T_c}}{T}}-1}\right)\,,
\end{equation}
and the $d_{x^2-y^2}$-pairing symmetry in the resulting gap
$\Delta ({\bf p},T)$ enters as follows
\begin{equation}
\Delta ({\bf p},T) = \Delta (T)\left[ \frac{\cos (p_xa/\hbar
)-\cos (p_ya/\hbar
)}{%
1-\cos (p_Fa/\hbar )}\right]\,.
\end{equation}

We showed recently\cite{oca3} that at low temperatures and high
fields, a quantitative agreement between this model and the
oscillation amplitudes of the thermal conductivity tensor is
achieved only if the intervortex spacing is increased by about
five times the value defined in Eq.(\ref{spacing}) if we use the
parameters from Ref.~\cite{yu} ($\Delta_0 = 20$~meV,
Ginzburg-Landau parameter $\kappa = 100$, anisotropy $\gamma = 4$,
$B^{ab}_{c2} = 650$~T). Furthermore, from those fits\cite{oca3}
for the twinned crystal we obtained a momentum independent
impurity scattering $\Gamma^{-1}_{\rm imp} \simeq 0.12$~ps and
$(\Gamma^{0}_{v})^{-1}(B = B_{c2}) \simeq (3/2)\Gamma^{-1}_{\rm
imp}$. Assuming a total scattering rate given by the sum of the AS
and impurity rates, i.e. $\Gamma ({\bf B},{\bf p},T) =
\Gamma_v({\bf B},{\bf p},T) + \Gamma_{\rm imp}$, the model given
by Eqs.~(1) to (5) reproduces remarkably well the symmetry of the
curves in Fig.~\ref{fig4} in the whole measured temperature range.
Because in this calculation we do not take explicitly into account
the inelastic scattering rate $\Gamma_{\rm in}$ given by phonons
$\Gamma_{\rm ph}$ and by quasiparticles $\Gamma_{\rm qp}$,  the
calculated oscillation amplitude above 20~K increases up to $\sim
10$ times that observed in the experiment.

\begin{figure}
\vspace{-1.50cm}
\centerline{\psfig{file=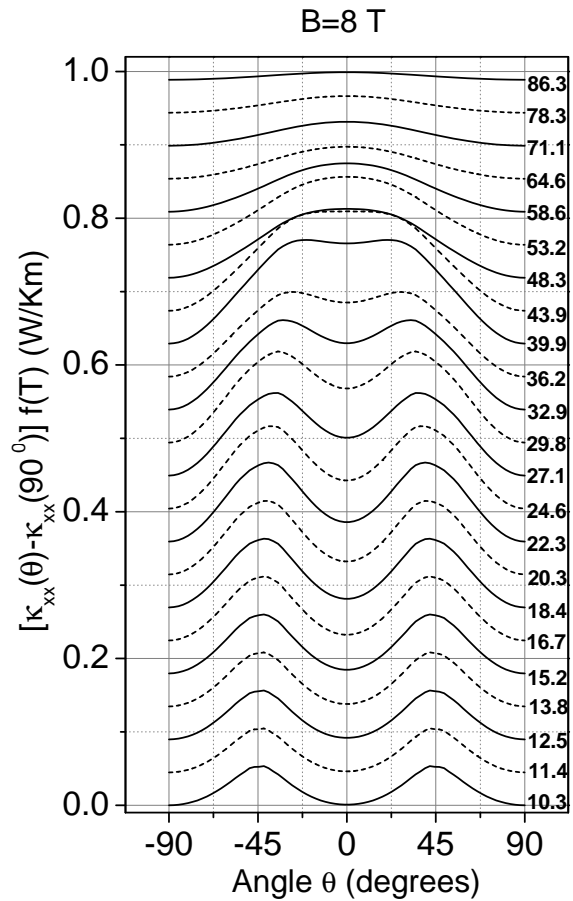,width=\columnwidth}}
\caption[*]{Angle dependence patterns of the longitudinal thermal
conductivity at different temperatures and 8 T calculated using
Eqs.~(1) to (5) and with parameters given in the text. Each of the
curves at a given temperature has been multiplied by a free
parameter $f(T)$ in order to fit the experimental oscillation
amplitude of the twinned sample. An offset has been added to every
curve in order to distinguish the change developed in the symmetry
as the temperature increases.} \label{fig5}
\end{figure}

In fig.~\ref{fig5} we show the results of the numerical
simulation. Each of the curves in this figure has been multiplied
by a factor $f(T)$ in order to fit the experimental oscillation
amplitude (Fig.~\ref{fig4}). We note that the symmetry of the
results can be explained quite satisfactorily by this model.
This result, on one hand, supports the predominance of
$d_{x^2-y^2}$-pairing symmetry of the order parameter up to
temperatures close to $T_c$ and, on the other hand, confirms the
idea of a competition between the thermal activation of
quasiparticles in the direction of the thermal current and the gap
structure. We note that the $d_{x^2-y^2}$-gap symmetry is the only
gap function that can explain the whole angular patterns up to
T$_c$. Of course, a $s$-gap also gives a onefold oscillation
similar to the experiments above $\sim$~50~K, but it does produce
neither the fourfold oscillation below $\sim$~15~K nor the curves
between 15~K and 50~K. Thus, the set of curves in Fig.~\ref{fig4}
can be only explained if one uses a main $d_{x^2-y^2}$-gap into
the calculations. Furthermore , the anisotropy of $d$-wave gap
proposed for BSCCO,\cite{ando} does not seem to occur in YBCO
since in the latter the nodes are observed at both field
orientations $\theta$=~45$^\circ$ and $\theta$=~-45$^\circ$(see
also Refs. \cite{aub,yu,oca3}).

\begin{figure}
\begin{center}
\centerline{\psfig{file=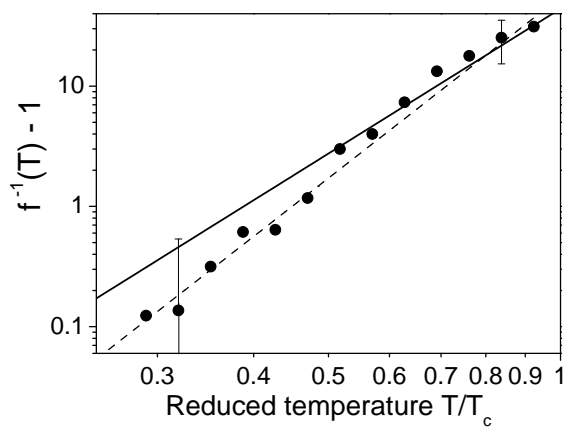,width=\columnwidth}}
\end{center}
\caption[*]{Temperature dependence of the factor $f^{-1}(T)-1$ as
a function of reduced temperature in a logarithmic scale. The
dashed line is a fit of the data to a $T^5$ dependence and the
solid line follows a $T^4$ dependence.} \label{fig6}
\end{figure}

The factor $f(T)$ is related to the inelastic scattering rate
$\Gamma_{\rm in}$, which was not taken explicitly into account in
the previous calculations. However, an approximation can be
carried out to get roughly the temperature dependence of the total
scattering rate, which could be in part associated to the
temperature dependence of $\Gamma_{\rm in}$. Since the AS
mechanism is weakly temperature dependent below $T/T_c < 0.8$, we
may approximate the total scattering rate as
\begin{equation}
\Gamma_{\rm imp}+\Gamma_{v}+\Gamma_{\rm in} \sim (\Gamma_{\rm imp}
+ \Gamma_{v})f^{-1}(T)\,. \label{aprox}
\end{equation}
The temperature dependence of the inelastic scattering rate is
given then by the factor $f(T)$ in this approximation as
\begin{equation}
{\Gamma_{\rm in} \propto f^{-1}(T)-1 }\,. \label{aprox2}
\end{equation}
In Fig.~\ref{fig6} we plot $f^{-1}(T)-1$.
As expected, we recover the overall behavior of the inelastic
scattering rate already described by many
authors\cite{hir2,quin,houssa1,houssa2,houssa3}. Below $T_c$ it
decreases rapidly and becomes negligible in comparison with the
impurity and AS scattering rates below $\sim 25$~K. Although the
approximation given by (\ref{aprox}) is too rough to obtain the
true temperature dependence of the inelastic scattering rate, it
is instructive to compare the obtained power dependence with
results from literature. Thermal Hall angle measurements performed
in the same YBCO twinned crystal show that $\cot(\theta_H) =
m_H/\tau_H \propto T^4$ down to $\sim 20~$K ($m_H$ is the
effective mass and $\tau_H$ is the Hall scattering time of the
quasiparticles responsible for the Hall signal).\cite{oca1} The
power law dependence obtained for $\tau^{-1}_H \propto T^4$ is
similar to that obtained for the longitudinal scattering rate
assuming a $d-$wave pairing.\cite{yu2} Since
quasiparticle-quasiparticle scattering mechanism should be the
dominant temperature dependent inelastic scattering below $T_c$,
we expect a rate proportional to the density of quasiparticles.
Interestingly, within the simple two-fluid model we expect a
density of quasiparticles proportional to $(T/T_c)^4$. On the
other hand, a $T^3$ dependence is expected within the
spin-fluctuation scattering picture\cite{quin}.

\section{Summary}\label{sum}
In summary, we have measured the longitudinal thermal conductivity
$\kappa_{xx}$ in two single crystals of YBCO in the presence of a
planar magnetic field which was rotated parallel to the CuO$_2$
planes,  from $T \sim 10$ K up to a few Kelvins below $T_c$.
Fourfold oscillations were recorded below $\sim 20$ K. Above this
temperature the angle dependence of the longitudinal thermal
conductivity changes; from the minimum at $\theta = 0^\circ$
(field parallel to the heat current) a maximum develops at high
$T$. The observed behavior in the whole temperature range can be
very well reproduced  by a model involving Andreev scattering of
quasiparticles by vortices and the two-dimensional BRT expression
for the thermal conductivity assuming a $d-$wave pairing. The
overall results agree with the $d_{x^2-y^2}$-pairing symmetry of
the order parameter. This agreement suggests that the mechanisms
that influence the behavior of the quasiparticles below $T_c$ and
above $\sim 10~$K are well described by the Fermi liquid theory at
nearly optimal doping. The small differences found in the angle
patters between the twinned and untwinned samples can be explained
in terms of the different impurity concentration as well as oxygen
deficiency and are accounted for by the phenomenology exposed in
this work.

\begin{acknowledgments}
This work was supported by the DFG under Grant DFG ES 86/4-3.
\end{acknowledgments}

\end{document}